# Emerging interfacial magnetization in isovalent manganite heterostructures driven by octahedral coupling

Yogesh Kumar[1,@], Harsh Bhatt[1,2], S. Kakkar[3], C. J. Kinane[4], A. Caruana[4], S. Langridge[4], Chandan Bera[3], S. Basu[1,2], Manuel A. Roldan[5] and Surendra Singh[1,2,*]

[1]Solid State Physics Division, Bhabha Atomic Research Centre, Mumbai 400085 India

[2]Homi Bhabha National Institute, Anushaktinagar, Mumbai 400094 India

[3] Quantum Materials and Devices Unit, Institute of Nano Science and Technology, Phase- 10, Sector- 64, Mohali, Punjab - 160062, India

[4]ISIS-Neutron and Muon Source, Rutherford Appleton Laboratory, Didcot, Oxon OX11 0QX, United Kingdom

[5]Eyring Materials Center, Arizona State University, Arizona 85287, USA

[@]Present Address: UGC-DAE Consortium for Scientific Research, 246-C CFB, BARC, Mumbai 400085, India

*surendra@barc.gov.in

The distortion of corner-sharing octahedra in isovalent perovskite transition-metal oxide interfaces is proven to be an excellent way to tailor the electronic and magnetic properties of their heterostructures. Combining depth-dependent magnetic characterization technique; (polarized neutron reflectivity, PNR); and theoretical calculation (density functional theory), we report interface-driven magnetic exchange interactions due to a modification in the octahedral rotations at the interfaces in an isovalent $La_{0.67}Ca_{0.33}MnO_3$ (LCMO)/$La_{0.67}Sr_{0.33}MnO_3$ (LSMO) heterostructures. PNR results determined a length scale of ~ 8 unit cells at the interface, which demonstrated a modification in magnetic properties. The results also predicted a low-temperature exchange bias for these ferromagnetic heterostructures with a maximum exchange bias for the heterostructure, which showed an enhanced antiferromagnetic coupling at the interfaces.

The emergent interfacial phenomena in transition metal perovskite oxide ($ABO_3$) heterostructures are induced by the strong coupling between spin, orbital, charge and lattice degrees of freedom [1,2]. The interfacial phenomena with intriguing physics in oxide heterostructures have been investigated mostly by considering the charge transfer and structural proximity/coupling at the heterointerfaces [1,3,4]. The interfacial charge transfer/redistribution as a result of the alignment of bands at the heterointerfaces [5-8], can intrinsically cause hole/electron doping without inducing chemical disorder and contributes to the key emerging interfacial properties in heterostructures [9,10]. The $BO_6$ oxygen octahedra, which are intimately correlated to orbital, charge and spin order in perovskite oxides, enable the structural distortion by decreasing the B-O-B bond angles ($\beta$) and increasing the B-O bond lengths ($d$). These structural modifications lead to a decrease in the electronic bandwidth ($W$) as $W \propto \cos[0.5\,(\pi - <\beta>)]/d^{3.5}$ [11], and directly change the electronic and magnetic properties. In the case of $ABO_3$ heterostructures, which offer additional means to tune the lattice structure, the oxygen octahedral rotation can be regulated either by interfacial strain or by interfacial oxygen octahedral coupling (OOC) [12-17]. The OOC is a geometric constraint effect that can alter the amplitude of octahedral rotations over roughly < 10 unit cells (*u.c.*) on either side of a heterointerface [12-17]. Therefore, studies are required to explore the magnetic modulation on this length scale at the heterointerfaces.

The purely structural proximity at interfaces has been known as a $\delta$-doping strategy, which controls the OOC and provides enhanced magnetization at the interfaces in manganite heterostructures [18-20]. The $\delta$-doping strategy in manganite heterostructures was achieved by inserting an atomically thin manganite layer into an isovalent manganite host, which modifies the local rotations of corner-connected $MnO_6$ octahedra [20]. Santos *et al.*, [18] observed an enhanced ferromagnetic (FM) exchange in otherwise antiferromagnetic (AFM) manganite $LaMnO_3/SrMnO_3$ superlattices due to $\delta$-doping by growing alternate single *u.c.* layer of these manganites. The ability to spatially confine magnetic states without altering the local charge density by local control of octahedral rotations as a $\delta$-doping approach was also achieved by growing ultrathin isovalent manganite superlattices of $La_{0.7}Sr_{0.3}MnO_3/Eu_{0.7}Sr_{0.3}MnO_3$ [19] and $La_{0.5}Sr_{0.5}MnO_3/La_{0.5}Ca_{0.5}MnO_3$ [20]. Here we show experimental evidence of interface-driven enhancement as well as suppression of ferromagnetism at interfacial $La_{0.67}Ca_{0.33}MnO_3$ (LCMO) region in $La_{0.67}Sr_{0.33}MnO_3$ (LSMO)/LCMO heterostructures with different stacking sequence using polarized neutron reflectivity (PNR) experiments. PNR results suggest a finite magnetic moment for an interfacial LCMO layer of thickness ~ 8 *u.c.* at the LCMO/LSMO interface above the FM transition temperature of the LCMO layer. Moreover, we found a shift in the hysteresis loop along the field axis (exchange bias) at low temperatures. Our theoretical investigation shows different values of nearest neighbour exchange interactions due to a local rotation pattern of the octahedra, which is instrumental in stabilizing these interface-dependent magnetic interactions and couplings in these heterostructures.

Isovalent manganite heterostructures of LSMO/LCMO with different thicknesses and stacking sequences of the LSMO and LCMO layers were grown on MgO (001) substrates (Table 1) using pulsed laser deposition (PLD) technique. Different thicknesses and stacking sequences of the manganite layers were used to study the effects of strain and interface coupling in these systems. The structural details of the bilayer heterostructures S1 (LSMO/LCMO/MgO) and S2 (LCMO/LSMO/MgO), as well as multilayer S3 ([LSMO/LCMO]$_5$/MgO), are given in Table 1 and the schematic of these heterostructures is shown in the inset of Fig 1(a). A KrF excimer laser (wavelength = 248 nm, pulse width = 20 ns) was used to ablate the high-purity bulk targets with a laser energy density of ~ 4 J/cm$^2$. The base pressure of the PLD chamber was reduced to 10$^{-6}$ mbar before the deposition. The films were deposited at an oxygen partial pressure of 0.2 mbar and the substrate temperature was kept at 750 °C.

The x-ray diffraction (XRD) scans for different heterostructures as well as a single layer of LSMO on the MgO (001) substrate are shown in Fig. 1(a). A comparison of XRD data from different heterostructures suggests crystalline growth with the (001) texture of manganite films. The high intensity, (002) Bragg peak for manganite films in different heterostructures is used to estimate the strain in the films, shown in Fig. 1(b), suggesting that films are relaxed. The MgO and LSMO/LCMO have a lattice constant of ~ 4.2 Å and 3.87 Å (shown by horizontal lines in Fig. 1 (b)), respectively, in their bulk phase, suggesting a tensile lattice mismatch strain ($\varepsilon$) of ~ 8.0%. However, the lattice constant of manganite films in these heterostructures was found to almost matching to its bulk value (i.e. $\varepsilon$ ~ 0.1% to 0.7%, see Table 1), which is consistent with an earlier study [21], indicating that oxide film grown on MgO substrate relaxes fast with respect to the thickness of the film (~ 50 Å). The depth-dependent chemical structures (thickness, interface roughness and electron scattering length density (ESLD)) of these heterostructures have been studied using x-ray reflectivity (XRR) measurements (Fig. S1 in Supplemental Material [22]) and the structural parameters are given in Table 1. XRR measurements suggested a high-quality layer structure with small interface roughness (~ 5 Å). Chemical depth profiles obtained for different heterostructures from XRR data were subsequently used to fit the PNR data by varying the parameters within a constrained range.

Further, the atomic-scale structure of the heterostructure was investigated by atomic resolution electron microscopic studies using aberration-corrected scanning transmission electron microscope (JEOL-ARM) to acquire high-angle annular dark field (HAADF) images. A representative cross-section HAADF image of heterostructure S3 (a multilayer) is shown in Fig. 1(c), showing the high-quality epitaxial growth of the multilayer. In addition, energy dispersive x-ray spectroscopy (EDS) measurements were performed to distinguish the LCMO and LSMO layers. The elemental maps determined from characteristic La-$L_{\alpha 1}$, O-$K_{\alpha 1}$, Mn-$K_{\alpha 1}$, Ca-$K_{\alpha 1}$ and Sr-$K_{\alpha 1}$ EDS edges from the heterostructure S3 (Fig. S2 in Supplemental Material [22]) suggest high-quality layer structure. The depth profiling of elemental concentration using scanning transmission electron microscopy (STEM), shown in Figs. 1(d) and (e), and the STEM image of heterostructure S3 suggest a well-defined multilayer structure with uniform distribution of La, O and Mn elements and well-defined Ca and Sr-rich regions (layers) for S3. Therefore, combining scattering techniques (x-ray and neutron) with a direct imaging technique (STEM) for the structural characterization of LSMO/LCMO heterostructures clearly suggested high-quality interfaces.

Figure 1 (f) shows the zero-field-cooled (ZFC) and field-cooled (FC) (cooling field, $H_{FC}$ = 500 Oe) magnetization of multilayer S3 as a function of temperature, $M(T)$, measured in an in-plane applied field, $H_{FC}$ = 500 Oe, using a Quantum Design superconducting quantum interference device (SQUID) magnetometer MPMS5. The $M(T)$ data for S3 (Fig. 1(f)) and other heterostructures (Fig. S3 in Supplemental Material [22]) indicate that the Curie temperature ($T_C$) for these heterostructures is lower than 300 K and is given in Table 1. The $M(T)$ data for similarly grown single LSMO and LCMO films of thickness ~ 100 Å on MgO substrates are also shown in Fig. 1 (g), suggesting a $T_c$ of ~ 295 K and 140 K for LSMO and LCMO films, respectively, which is consistent with results on similar systems [23, 24]. The smaller $T_c$ in these heterostructures agrees with previous studies [23-27], which is attributed to deposition conditions (oxygen pressure), different thicknesses, and charge discontinuity in manganite films. Typical hysteresis loops ($M$ (H) curve) at different temperatures (200, 100 and 5 K) for S3 measured under FC ($H_{FC}$ = 500 Oe) condition are shown in Fig. 1(h). Interestingly, we find a small shift in the hysteresis curve (exchange bias) to the negative field (Fig. 1 (h)) at low temperatures for all the heterostructures. $M$ (H) hysteresis curves at 5 K for single LSMO and LCMO films grown on MgO substrates, measured under identical conditions as that of heterostructures (i.e. FC with $H_{FC}$ = 500 Oe), are shown in Fig. 1 (i), which are symmetric about field axis. The exchange bias field ($H_{EB}$), coercive field ($H_c$) and average magnetization at 5 K for different heterostructures, shown in Fig. 1 (j), indicate a maximum negative $H_{EB}$ of ~ 60 Oe for heterostructure S2. The existence of small but finite $H_{EB}$ for LCMO/LSMO heterostructures, even though both LSMO and LCMO are FM at low temperatures, clearly indicates a magnetic modification at the interfaces.

To investigate the interfacial coupling in these isovalent manganite heterostructures, we carried out spin-dependent PNR [28-31] measurements for all the heterostructures at different temperatures (300, 200, 75 and 10 K) across the $T_c$ of the LCMO film (~ 140 K). PNR measurements were carried out using the POLREF instrument at ISIS neutron and muon source, RAL, UK. PNR measurements were performed under FC ($H_{FC}$ = 500 Oe) conditions in the applied in-plane field of 500 Oe. Spin-dependent PNR, $R^+$ (spin up) and $R^-$ (spin down), where the +(−) sign denotes the neutron beam polarization parallel (opposite) to the applied field, providing both the nuclear and magnetic scattering length density (NSLD and MSLD) depth profiles of the heterostructures [28-32]. The schematic of the spin-dependent PNR measurements from a film is represented in the inset of Fig. 2 (a), where $H$ is the applied field in the heterostructures. The PNR results for heterostructures S1 and S2 are shown in Figs. 2 (a) and (b) respectively, where PNR data (symbols) and corresponding fits (solid lines) at different temperatures are shifted vertically for better visualization. The NSLD and magnetization ($M$ = MSLD/2.91×10$^{-9}$ [28]) depth profiles of the heterostructures were obtained by fitting the PNR data using an optimization program [28], which uses Parratt formalism [33] and the parameters are adjusted to minimize the value of reduced $\chi^2$ –a weighted measure of the goodness of fit. The difference between spin-dependent PNR data (i.e., $R^+$ - $R^-$) provides the detailed magnetization depth profile of the heterostructure. As expected, PNR data at 300 K show the negligible difference between $R^+$ and $R^-$ and are used to obtain the detailed NSLD depth profiles for the heterostructures S1 and S2, shown in Figs. 2 (c) and (e), respectively, which are consistent with the corresponding XRR results (see Fig. S1 in Supplemental Material [22]) from these heterostructures. The corresponding magnetization depth profiles obtained from PNR at different temperatures are shown in Figs. 2 (d) and (f). PNR results provided similar values of thickness and interface roughness parameters (within the error on the parameters) as obtained from the XRR (Table 1).

Remarkably, we find the emergence of a finite magnetization (FM order) for an interfacial LCMO layer of thickness ~ 32 Å (~ 8 *u.c.*) in S1 at a temperature of 200 K (shaded region in Fig. 2 (d)), which is much higher than the $T_c$ (~ 140 K) for a similarly grown single LCMO layer. While such an emergent interfacial magnetization for heterostructure S2 (Fig. 2(f)), which is grown with a reverse stacking sequence, was completely absent at 200 K. To validate the PNR results at 200 K for S1 and S2, we have fitted the PNR data considering different models of interface magnetization and compared them using normalized spin asymmetry (NSA) profiles. NSA profiles are the ratio of the difference and sum of the spin-dependent PNR data ($R^+$ and $R^-$), i.e., NSA = ($R^+$ - $R^-$)/ ($R^+$ + $R^-$). Figures 2 (g) and (h) show the NSA data (solid circles) for S1 and S2, respectively, at 200 K, which are fitted using different models with (solid red line) and without (black line with solid triangles) an interfacial FM LCMO layer (Figs. 2(i) and (j)). These models are statistically compared for the quality of fits using a goodness-of-fit parameter, $\chi^2$ ( $= \sum_i \left[\frac{NSA_{\exp}(Q_i) - NSA_{th}(Q_i)}{err(Q_i)}\right]^2$, where $NSA_{exp}(Q_i)$, $NSA_{th}(Q_i)$ and $err(Q_i)$ are the NSA data points, NSA for the fitted model and error on data points at wave-vector transfer $Q_i$, respectively). The comparison of different interfacial magnetization models (Figs. 2 (i) and (j)) confirms the emergence of ordered ferromagnetism for interfacial LCMO layer (thickness ~ 32 Å) for S1 at 200 K, which is well above the $T_c$ of LCMO film, and no interfacial ordered magnetization when the growth sequence of layers is reversed in heterostructure S2. It is noted that the thickness of the interfacial FM LCMO layer (~ 32 Å) at 200 K is much larger than the interface roughness (~ 5 Å). Moreover, reflectivity data also suggested a similar value of the roughness (~ 5 Å) for both LSMO/LCMO and LCMO/LSMO interfaces in these heterostructures. This rules out the possibility of asymmetric chemical structure at the interfaces as a plausible reason for the observed effect and thus it indicates the asymmetric magnetic nature is an intrinsic behaviour. To confirm such magnetic asymmetric behaviour in these isovalent manganite heterostructures, which depend on the sequential growth of layers, we have studied the depth-dependent magnetic properties in a multilayer (heterostructure S3) providing two different interfaces in a single heterostructure (Fig. 3).

Figure 3 depicts the PNR results from heterostructure S3. PNR data for S3 at different temperatures, which are vertically shifted for better visualization, are shown in Fig. 3 (a). NSLD and magnetization (at different temperatures) depth profiles for S3 obtained from PNR data are shown in Figs. 3(b) and (c), respectively. The temperature-dependent magnetization depth profiles (Fig. 3(c)) of heterostructure (multilayer) S3 suggest the emergence of an interfacial LCMO magnetic layer and asymmetric magnetic behaviour at structurally symmetric interfaces of S3. We find an ultrathin FM interfacial LCMO layer at LSMO/LCMO (LSMO grown on LCMO) interface and the absence of such interfacial ferromagnetism at LCMO/LSMO (LCMO grown on LSMO) interface at 200 K. This supports the asymmetric magnetic behaviour of interfacial LCMO layer at 200 K (well above the $T_c$ of LCMO film), observed in heterostructure S1 and S2. To confirm such asymmetric emergent FM behaviour at the interfaces of S3 at 200 K we have fitted PNR data (Figs. 3(d to g)) assuming different magnetization depth profiles (Figs. 3 (h to k)). Different magnetization depth profiles across two interfaces of S3 and the corresponding fits to NSA data at 200 K are indicated by vertical arrow (both sides) in the middle and lower panel of Fig. 3. Different interfacial magnetization models were compared with goodness-of-fit parameter ($\chi^2$) and it is indicated in the Figs. 3 (d to g). The mismatch between experimental NSA (or PNR) data at 200 K and fits assuming different interfacial magnetization models, especially around the Bragg peak position ($Q_z$ ~ 0.05 Å$^{-1}$), as compared to the best-fit model (Fig. 3 (g) and (k)) further confirms the emergence of FM for the interfacial LCMO layer at the LSMO/LCMO interface and the absence of FM for the interfacial LCMO layer at another (LCMO/LSMO) interface. In addition, we find enhanced and reduced magnetization for the interfacial LCMO layer at the LSMO/LCMO and the LCMO/LSMO interfaces, respectively, as compared to the rest of the LCMO layer in these heterostructures for all temperatures of measurements below $T_c$ of the LCMO film. Thus, the depth-dependent magnetization for different heterostructures at 200 K suggests the emergence of stacking sequence-dependent FM order for the interfacial LCMO layer, well above its transition temperature.

To understand the existence of this ultrathin interfacial FM LCMO layer at the LSMO/LCMO interface above $T_c$ and related asymmetric magnetic behaviour at interfaces, as well as observation of exchange bias in these all FM-based heterostructures using macroscopic magnetization measurements, we have performed first-principle calculations based on density functional theory (DFT) using the projector augmented-wave (PAW) method as implemented within the Vienna Ab-initio Simulation Package (VASP) [34-36]. A generalized gradient approximation (GGA) with Perdew-Burke-Ernzerhof (PBE) exchange-correlation functional was used [37]. To simulate the two interfaces, (i) for LCMO/LSMO interface: a stoichiometric thin film model having LCMO thin film (3 *u.c.*) on an LSMO substrate with vacuum on the top of LCMO was considered, and (ii) for LSMO/LCMO interface: LSMO thin film (3 *u.c.*) on LCMO substrate with vacuum on the top of LSMO was considered. To simplify and reduce the complex nature of the computational analysis we have considered only a few *u.c.* (~ 3) for the thin film. To avoid the interaction between the adjacent slabs, a vacuum region of 15 Å was placed between the simulated slabs. These calculations were performed with a kinetic energy cut-off of 520 eV for a plane-wave basis [36] and a gamma-centred 4 × 2 × 1 k-point grid. To account for the electronic correlation between the localized electrons of 3*d* orbitals of Mn, DFT+U calculations were performed using Dudarev's rotationally invariant approach [38] with $U$ = 4 eV, $J$ = 0.0 eV for the Mn atom, where $U$ is the effective on-site Coulomb interaction between localized 3*d* electrons and $J$ is the exchange parameter.

In ABO$_3$ perovskite manganite oxides, the local magnetic exchange interactions are enhanced within the spatially confined regions of the suppressed octahedral rotations (increase in B-O-B bond angles) [20]. The depth-dependent octahedral rotation optimized structures of LCMO/LSMO and LSMO/LCMO heterostructure slabs obtained from the theoretical calculations are shown in Fig. 4(a) and (c), respectively. The in-plane ($\theta_{in}$) and out-of-plane ($\theta_{out}$) Mn-O-Mn bond angles are calculated for the optimized structures and are also shown in Fig. 4 (a) and (c) for two interfaces. For LCMO/LSMO system, the average values of $\theta_{in}$ for LCMO, interfacial region, and LSMO region are 156.8°, 161.05°, and 166.57°, respectively. Whereas, $\theta_{out}$ for LCMO, interface, and LSMO regions is 164.9°, 162.9°, and 174.42°, respectively. Thus, the bond angle order for LCMO/LSMO systems is LSMO > interface > LCMO for in-plane and LSMO > LCMO > interface for out-of-plane bond angle. As a result, $\theta_{out}$ governs the experimentally determined magnetization ordering in LCMO/LSMO system. For LSMO/LCMO system, the average values of $\theta_{in}$ for LSMO, interface, and LCMO regions are 168.95°, 165.05°, and 160.75°, respectively. In addition, $\theta_{out}$ for LSMO, interface, and LCMO regions is 168.5°, 165.8°, and 158.6°, respectively. Consequently, the bond angle order is LSMO > interface > LCMO for both $\theta_{in}$ and $\theta_{out}$. The higher values of Mn-O-Mn result in the increase in double exchange interaction responsible for magnetism in these systems. The bond order at two interfaces is consistent with the experimental magnetization ordering in LSMO/LCMO system. Furthermore, spin-polarized calculations provide information about the local magnetic moment on Mn atoms derived from the *d*-orbitals. The layer-wise calculated magnetic moment per Mn ion for the two heterostructures is depicted in Fig. 4 (b), which indicates the reduced and enhanced magnetization of interfacial LCMO layer at LCMO/LSMO and LSMO/LCMO interfaces, respectively. This indicates that the structural combination plays an important role in the magnetic properties of these perovskite manganite oxides. In addition, using the isotropic nearest-neighbour (NN) exchange interaction ($J_{12}^{interface}$) [39] between a pair of Mn ions (Mn1-O-Mn2) present at the interfaces, we found FM exchange interaction for the LSMO region at both the interfaces (LCMO/LSMO, and LSMO/LCMO). While the LCMO region showed FM and AFM exchange interaction at LSMO/LCMO and LCMO/LSMO interfaces, respectively. This finding of interface-driven FM and AFM exchange coupling for interfacial LCMO layer in these isovalent manganite heterostructures provides an excellent explanation for the experimental results obtained from PNR. The AFM exchange interaction for the interfacial LCMO layer in the case of heterostructure S2 (LCMO/LSMO interface) can be one of the reasons for observing the maximum exchange bias for this heterostructure at low temperatures.

PNR results from three different heterostructures (S1, S2 and S3) of LSMO-LCMO based system confirmed the evolution of asymmetric interfacial ferromagnetism in the LCMO layer at the interfaces well above the $T_c$ (~ 140 K). Depth-dependent structure of the heterostructures studied by reflectivity (both PNR and XRR) and EDS suggested intermixing at the interfaces, which can influence the magnetization at the interfaces. However, the length scale involved in interfacial magnetic modulation is found to be ~ 32 Å, which is about ~ 600% of the interface roughness (intermixing) observed for these heterostructures. Further, the DFT calculations suggested different octahedral rotations at interfaces, which modify the AFM and FM exchange interactions at different interfaces and thus exhibiting different and asymmetric magnetization as well as exchange bias in these systems.

In summary, we demonstrate the interface-driven coupling which controls the magnetic properties of isovalent correlated manganite heterostructures and exhibits exchange bias at low temperatures. The spin-dependent PNR technique enables us to determine a length scale of ~ 8 u.c. for the interfacial LCMO region, which exhibits either enhanced or reduced moment depending on the stacking-dependent interface coupling. The PNR results also suggest an FM behaviour of the interfacial LCMO region at the LSMO/LCMO interface above the transition temperature of LCMO. Furthermore, the DFT calculations reveal that the difference in magnetization at two interfaces (LCMO/LSMO and LSMO/LCMO) in these isovalent manganite heterostructures is due to the different magnitude of nearest-neighbour exchange interactions as a result of interface-driven deformation of oxygen octahedral at the interfaces.

**Acknowledgements**

We gratefully acknowledge late Dr C L Prajapat for SQUID measurements. This work is supported by the Department of Science and Technology (DST), India via the DST INSPIRE faculty research grant (DST/INSPIRE/04/2015/002938). Y.K. acknowledges the Science and Engineering Research Board (SERB), India for the financial support via research grant (No. SB/SRS/2021-22/65/PS). We thank the ISIS Neutron and Muon source for the provision of beam time (RB1868012; https://doi.org/10.5286/ISIS.E.RB1868012) and the DST, India (SR/NM/Z-07/2015) for the financial support for performing the experiment and Jawaharlal Nehru Centre for Advanced Scientific Research (JNCASR) for managing the project (SR/NM/Z-07/2015). We also thank Dr R. J. Choudhary and Dr V. R. Reddy from UGC-DAE,

CSR, Indore for sample preparation using pulsed laser deposition and x-ray reflectivity measurements, respectively. We acknowledge the use of facilities within the Eyring Material Center at Arizona State University.

Table 1: The LSMO/LCMO heterostructures used in this study. The thicknesses of different layers and interfacial roughnesses were obtained from XRR.

| Samples | Thickness (Å) | | Roughness (Å) | | Strain (%) | $T_c$ (K) |
|---|---|---|---|---|---|---|
| | LCMO | LSMO | LSMO/LCMO | LCMO/LSMO | | |
| S1: LSMO/LCMO/MgO | 104±7 | 52±4 | 6±2 | - | -0.70 | 265 |
| S2: LCMO/LSMO/MgO | 97±6 | 96±7 | - | 4±1 | +0.26 | 275 |
| S3: [LSMO/LCMO]$_5$/MgO | 98±7 | 53±5 | 5±2 | 4±1 | +0.13 | 265 |

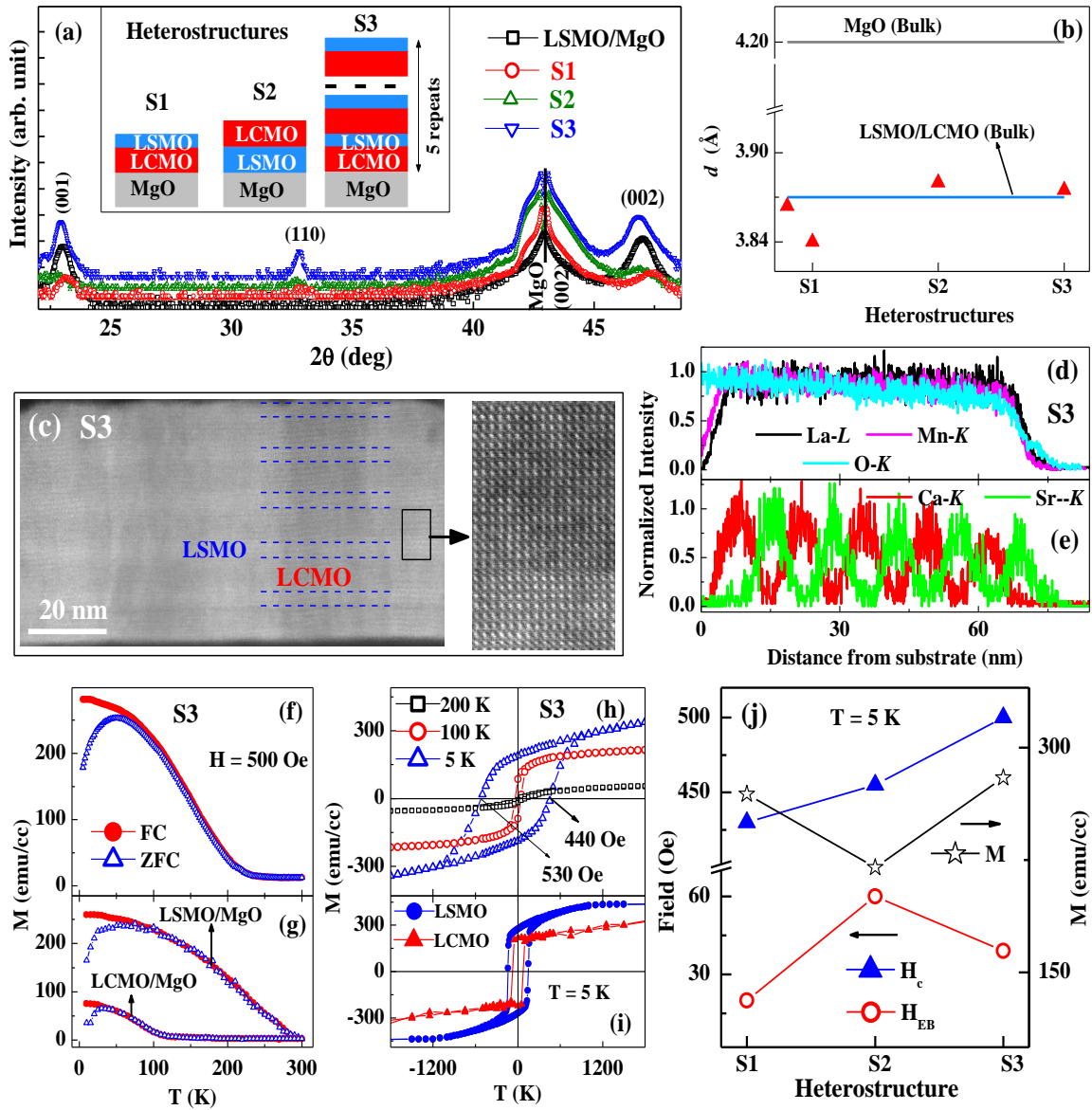

Figure 1: (a) XRD patterns for different LCMO/LSMO heterostructures grown on single-crystal MgO substrate. The inset shows the schematic of different heterostructures, S1, S2, and S3. (b) Lattice constant (*d*) measured from XRD for different heterostructures. (c) High-angle annular dark field image of heterostructure S3 measured by high-resolution STEM. (d and e) The depth profiling of the concentration of different elements for S3 is obtained from EDS measurements. (f) $M(T)$ measurements for S3 multilayer in an in-plane applied field of 500 Oe for field cooled (FC) and zero fields cooled (ZFC) conditions. (g) $M(T)$ data for single LSMO and LCMO layers on MgO substrates. $M(H)$ curves for (h) the S3 multilayer at different temperatures and (i) for single LSMO and LCMO films at 5 K. (j) Comparison of macroscopic magnetic properties ($H_{EB}$, $H_c$, average magnetization) at 5 K and variation of strain for different heterostructures.

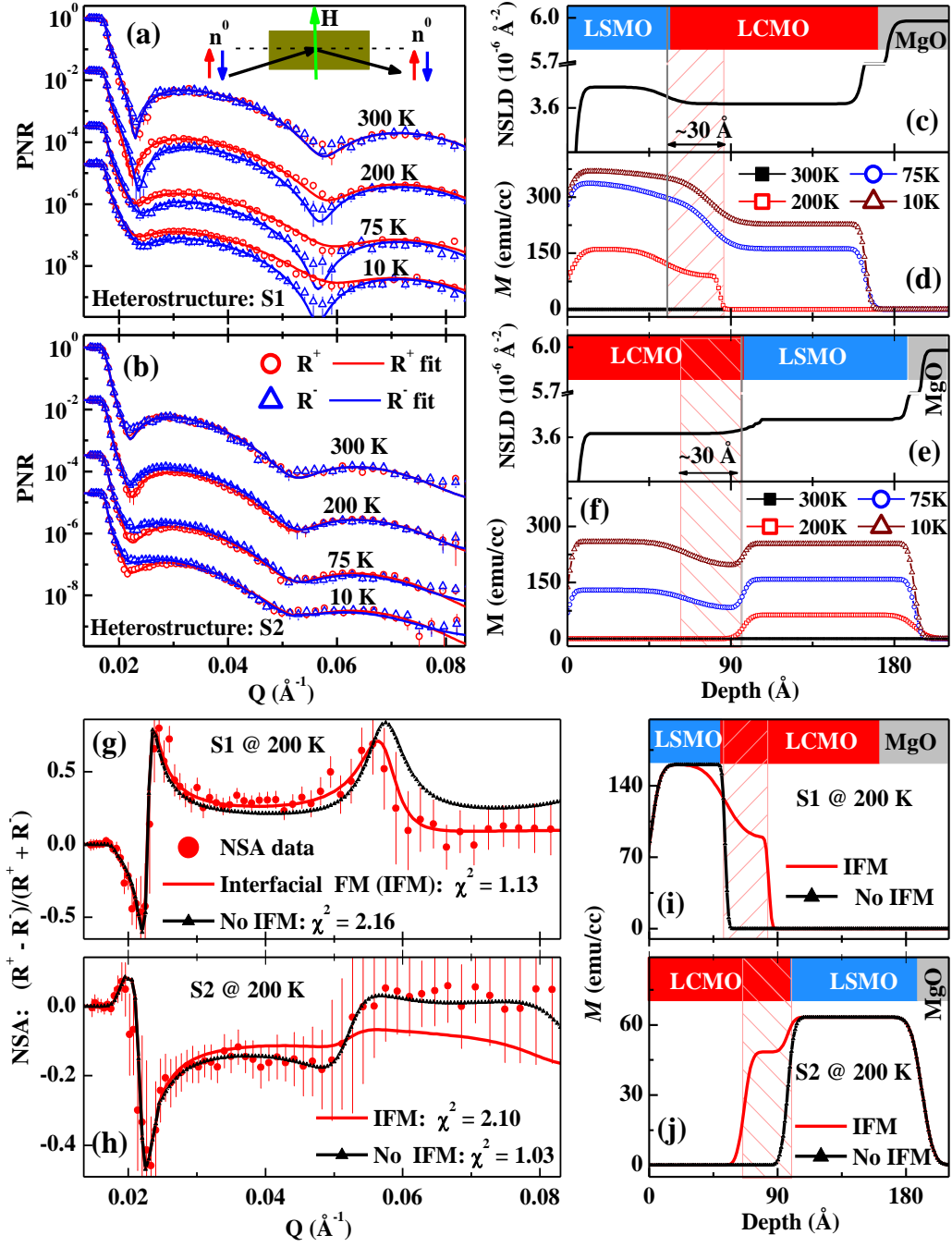

Fig. 2: Spin-dependent PNR data (symbols) and corresponding fits for (a) S1, and (b) S2 heterostructures at different temperatures, which are shifted vertically for better visualization. Inset shows the schematic of PNR from the film under an applied field of *H*. The nuclear scattering length density (NSLD) depth profile for (c) S1 and (e) S2 obtained from PNR data at 300 K. Magnetization (*M*) depth profiles at different temperatures for (d) S1, and (f) S2 extracted from the PNR data. The comparison of the normalised spin asymmetry (NSA) data for (g) S1 and (h) S2 heterostructures at 200 K using the magnetization models represented in (i) and (j) respectively.

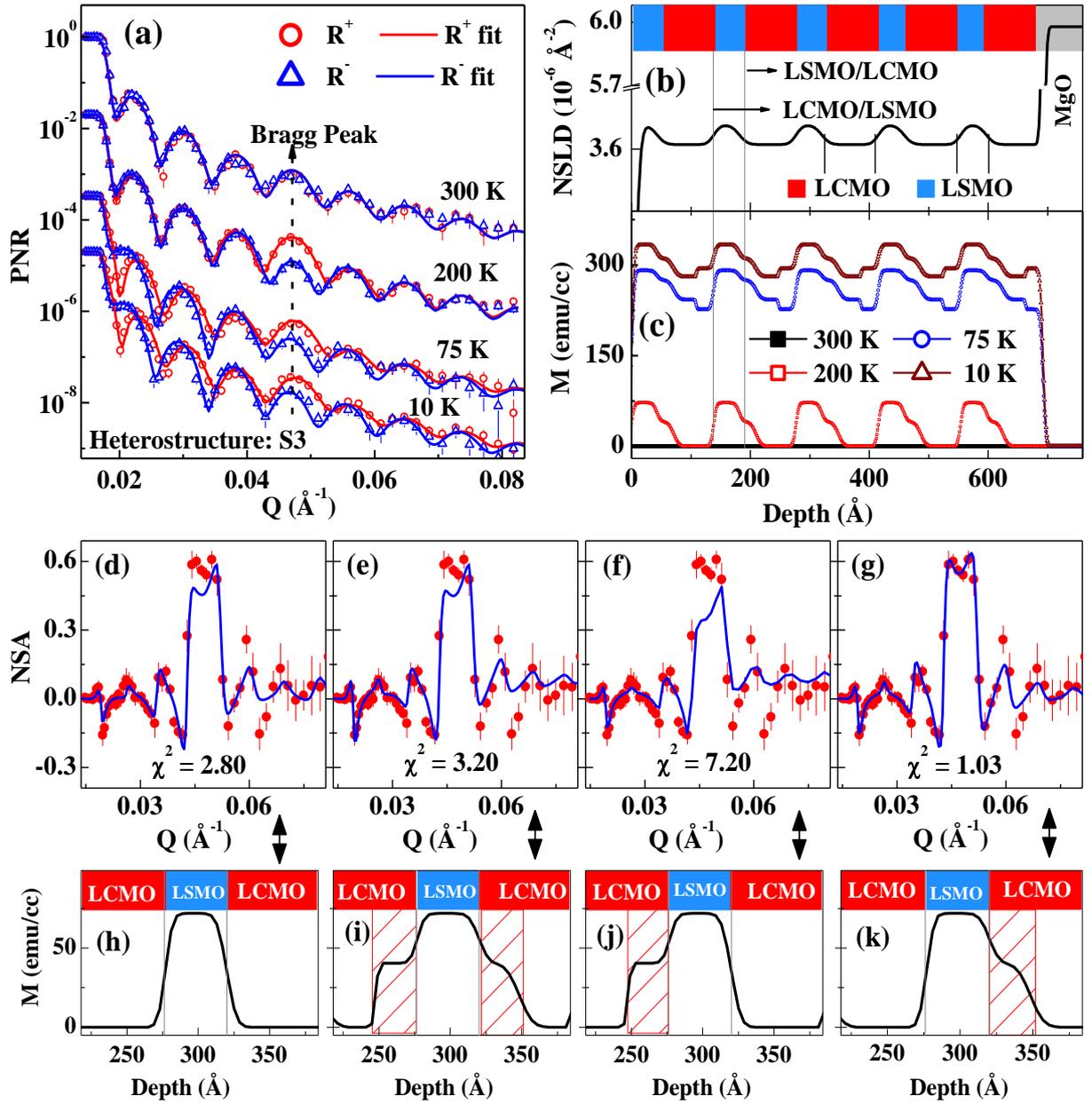

Figure 3: (a) Spin-dependent PNR data (symbols) and corresponding fits for heterostructure S3 at different temperatures, which are shifted vertically for better visualization. (b) The nuclear scattering length density (NSLD) depth profile for S3 extracted from PNR data at 300 K. (c) Magnetization (*M*) depth profiles at different temperatures for S3 extracted from the PNR data. The comparison of different magnetization models (h to k) across two interfaces (LSMO/LCMO and LCMO/LSMO) and the corresponding fits (d to g) to normalised spin asymmetry (NSA) data at 200 K for S3. Magnetization models and the corresponding fits to data are indicated by a vertical arrow.

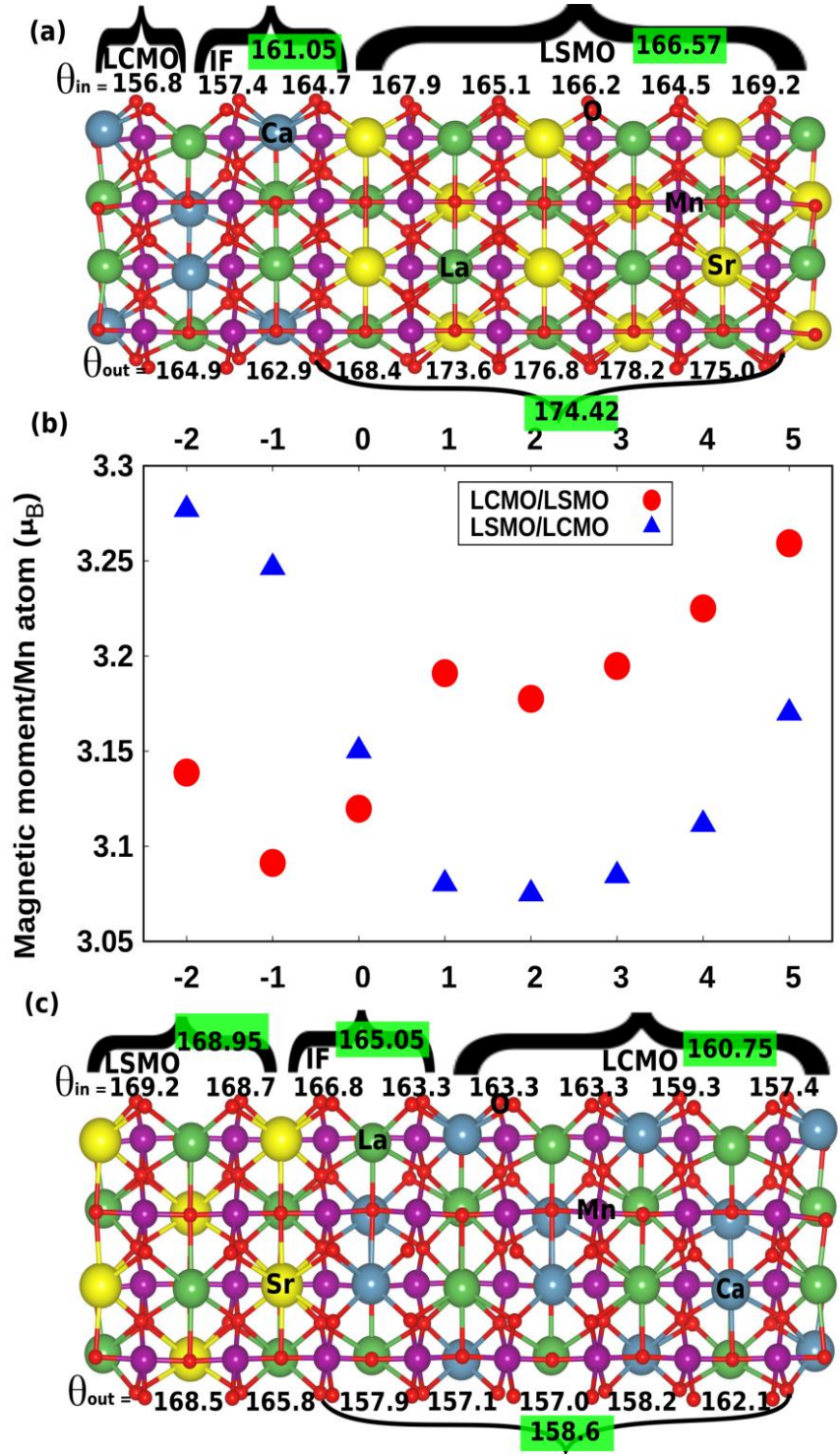

Figure 4: The relaxed crystal structures indicating the in-plane $\theta_{in}$ and out-of-plane $\theta_{out}$ bond angles of heterostructure (a) LCMO/LSMO and (c) LSMO/LCMO systems. The atoms O, La, Sr, Ca, and Mn are also marked on the structures. (b) The magnetic moment per Mn atom averaged over the number of atoms per layer vs. the layer number for LCMO/LSMO and LSMO/LCMO, respectively.

**References:**

[1] H. Y. Hwang, Y. Iwasa, M. Kawasaki, B. Keimer, N. Nagaosa, and Y. Tokura, Nature Materials 11, 103 (2012).

[2] E. Dagotto, Science 309, 257 (2005).

[3] J. M. Rondinelli and N. A. Spaldin, Advanced Materials 23, 3363 (2011).

[4] Z. Huang, Ariando, X. Renshaw Wang, A. Rusydi, J. Chen, H. Yang, and T. Venkatesan, Advanced Materials 30, 1802439 (2018).

[5] S. J. May et al., Nature Materials 8, 892 (2009).

[6] X. R. Wang et al., Science 349, 716 (2015).

[7] M. N. Grisolia et al., Nature Physics 12, 484 (2016).

[8] Z. Zhong and P. Hansmann, Physical Review X 7, 011023 (2017).

[9] A. Brinkman et al., Nature Materials 6, 493 (2007).

[10] A. Ohtomo and H. Y. Hwang, Nature 427, 423 (2004).

[11] C. Cui and T. A. Tyson, Applied Physics Letters 84, 942 (2004).

[12] J. M. Rondinelli, S. J. May, and J. W. Freeland, MRS Bulletin 37, 261 (2012).

[13] J. He, A. Borisevich, S. V. Kalinin, S. J. Pennycook, and S. T. Pantelides, Physical Review Letters 105, 227203 (2010).

[14] C. L. Jia, S. B. Mi, M. Faley, U. Poppe, J. Schubert, and K. Urban, Physical Review B 79, 081405 (2009).

[15] A. Y. Borisevich et al., Physical Review Letters 105, 087204 (2010).

[16] S. J. May, C. R. Smith, J. W. Kim, E. Karapetrova, A. Bhattacharya, and P. J. Ryan, Physical Review B 83, 153411 (2011).

[17] R. Aso, D. Kan, Y. Shimakawa, and H. Kurata, Scientific Reports 3, 2214 (2013).

[18] T. S. Santos et al., Physical Review Letters 107, 167202 (2011).

[19] E. J. Moon, R. Colby, Q. Wang, E. Karapetrova, C. M. Schlepütz, M. R. Fitzsimmons, and S. J. May, Nature Communications 5, 5710 (2014).

[20] E. J. Moon, Q. He, S. Ghosh, B. J. Kirby, S. T. Pantelides, A. Y. Borisevich, and S. J. May, Physical Review Letters 119, 197204 (2017).

[21] L. S.-J. Peng, X. X. Xi, B. H. Moeckly and S. P. Alpay, Appl. Phys. Lett. 83, 4592 (2003).

[22] See Supplemental Material for details of TEM from S3, XRR measurements, and SQUID results from S1 and S2.

[23] P. Lecoeur, A. Gupta, P. R. Duncombe, G. Q. Gong, and G. Xiao, Journal of Applied Physics 80, 513 (1996).

[24] H. Chou et al., Thin Solid Films 515, 2567 (2006).

[25] A. Tebano et al., Physical Review Letters 100, 137401 (2008).

[26] F. Yang, N. Kemik, M. D. Biegalski, H. M. Christen, E. Arenholz, and Y. Takamura, Applied Physics Letters 97, 092503 (2010).

[27] E.-J. Guo, T. Charlton, H. Ambaye, R. D. Desautels, H. N. Lee, and M. R. Fitzsimmons, ACS Applied Materials & Interfaces 9, 19307 (2017).

[28] S. Singh, M. Swain, and S. Basu, Progress in Materials Science 96, 1 (2018).

[29] S. Singh, M. R. Fitzsimmons, T. Lookman, J. D. Thompson, H. Jeen, A. Biswas, M. A. Roldan, and M. Varela, Physical Review Letters 108, 077207 (2012).

[30] H. Bhatt, Y. Kumar, C. L. Prajapat, C. J. Kinane, A. Caruana, S. Langridge, S. Basu, and S. Singh, Advanced Materials Interfaces 7, 2001172 (2020).

[31] S. Singh, J.T. Haraldsen, J. Xiong, E.M. Choi, P. Lu, D. Yi, X.-D. Wen, J. Liu, H. Wang, Z. Bi, Y. Pu, M.R. Fitzsimmons, J.L. MacManus-Driscoll, R. Ramesh, J.-X. Zhu, A.V. Balatsky, and Q.X. Jia, Physical Review Letters, 113, 047204 (2014).

[32] S. Basu and S. Singh, Neutron and X-ray Reflectometry: Emerging phenomena at heterostructure interfaces, IOP Publishing Bristol, UK (2022); https://doi.org/10.1088/978-0-7503-4695-5

[33] L. G. Parratt, Physical Review 95, 359 (1954).

[34] P. E. Blöchl, Physical Review B 50, 17953 (1994).

[35] G. Kresse and D. Joubert, Physical Review B 59, 1758 (1999).

[36] G. Kresse and J. Furthmüller, Physical Review B 54, 11169 (1996).

[37] J. P. Perdew, K. Burke, and M. Ernzerhof, Physical Review Letters 77, 3865 (1996).

[38] S. L. Dudarev, G. A. Botton, S. Y. Savrasov, C. J. Humphreys, and A. P. Sutton, Physical Review B 57, 1505 (1998).

[39] H. J. Xiang, E. J. Kan, S.-H. Wei, M. H. Whangbo, and X. G. Gong, Physical Review B 84, 224429 (2011).

# Supplementary Material

# Emerging interfacial magnetization in isovalent manganite heterostructures driven by octahedral coupling

Yogesh Kumar[1,@], Harsh Bhatt[1,2], S. Kakkar[3], C. J. Kinane[4], A. Caruana[4], S. Langridge[4], Chandan Bera[3], S. Basu[1,2], Manuel A. Roldan[5] and Surendra Singh[1,2,*]

[1]Solid State Physics Division, Bhabha Atomic Research Centre, Mumbai 400085 India
[2]Homi Bhabha National Institute, Anushaktinagar, Mumbai 400094 India
[3] Quantum Materials and Devices Unit, Institute of Nano Science and Technology, Phase- 10, Sector- 64, Mohali, Punjab - 160062, India
[4]ISIS-Neutron and Muon Source, Rutherford Appleton Laboratory, Didcot, Oxon OX11 0QX, United Kingdom
[5]Eyring Materials Center, Arizona State University, Arizona 85287, USA

[@]Present Address: UGC-DAE Consortium for Scientific Research, 246-C CFB, BARC, Mumbai 400085, India

*surendra@barc.gov.in

**XRR measurements:**

X-ray reflectivity (XRR) and Polarized neutron reflectivity (PNR) are two non-destructive complementary techniques used to obtain the depth profiles of chemical and magnetic structures in multilayer samples with a depth resolution of sub-nanometre length scale, averaged over the lateral dimension (~ 100 mm$^2$) of the sample [1-3]. The specular reflectivity (angle of incidence = angle of reflection) is related to the square of the Fourier transform of the depth-dependent (Z) scattering length density (SLD) profile $\rho(z)$ (normal to the film surface or along the Z-direction) [1-3]. For XRR, $\rho_x(z)$ is proportional to electron density and termed electron scattering length density (ESLD) whereas for PNR, $\rho(z)$ consists of nuclear SLD (NSLD) and magnetic SLD (MSLD) such that $\rho^{\pm}(z) = \rho_n(z) \pm \rho_m(z) = \rho_n(z) \pm CM(z)$, where $C = 2.9109 \times 10^{-9}$ Å$^{-2}$ cm$^3$/emu, and $M(z)$ is the magnetization (emu/cm$^3$) depth profile [1-3]. $\rho_n(z)$ and $\rho_m(z)$ are NSLD and MSLD, respectively. The sign +(-) is determined by the condition when the neutron beam polarization is parallel (opposite) to the in-plane magnetization of the sample and corresponds to reflectivities, $R^{\pm}$. Specular XRR and PNR data can be fitted using a genetic algorithm-based optimization program [1] which uses Parratt formalism [3]. For XRR, the parameters of a model include layer thickness, interface (or surface) roughness and ESLD. For PNR one can obtain NSLD and MSLD simultaneously.

Figures S1 (a), (b) and (c) show the XRR data (symbol) and corresponding fit (solid line) for heterostructure S1, S2 and S3 respectively. The electron scattering length density (ESLD) depth profiles for S1, S2 and S3, which best fitted XRR data are depicted in Figs 1 (d), (e) and (f), respectively. We obtained nice Keissig oscillations corresponding to the thicknesses of the layers even with a poor ESLD contrast at interfaces for these heterostructures. We also obtained a small Bragg peak due to 5 bilayers repeat for heterostructure S3 (multilayer) in Fig. S1 (c). The absence of 2$^{nd}$ order Bragg peak for XRR data reconfirms the bilayer thickness of ~ 150 Å for heterostructure S3 and a well-defined multilayer structure.

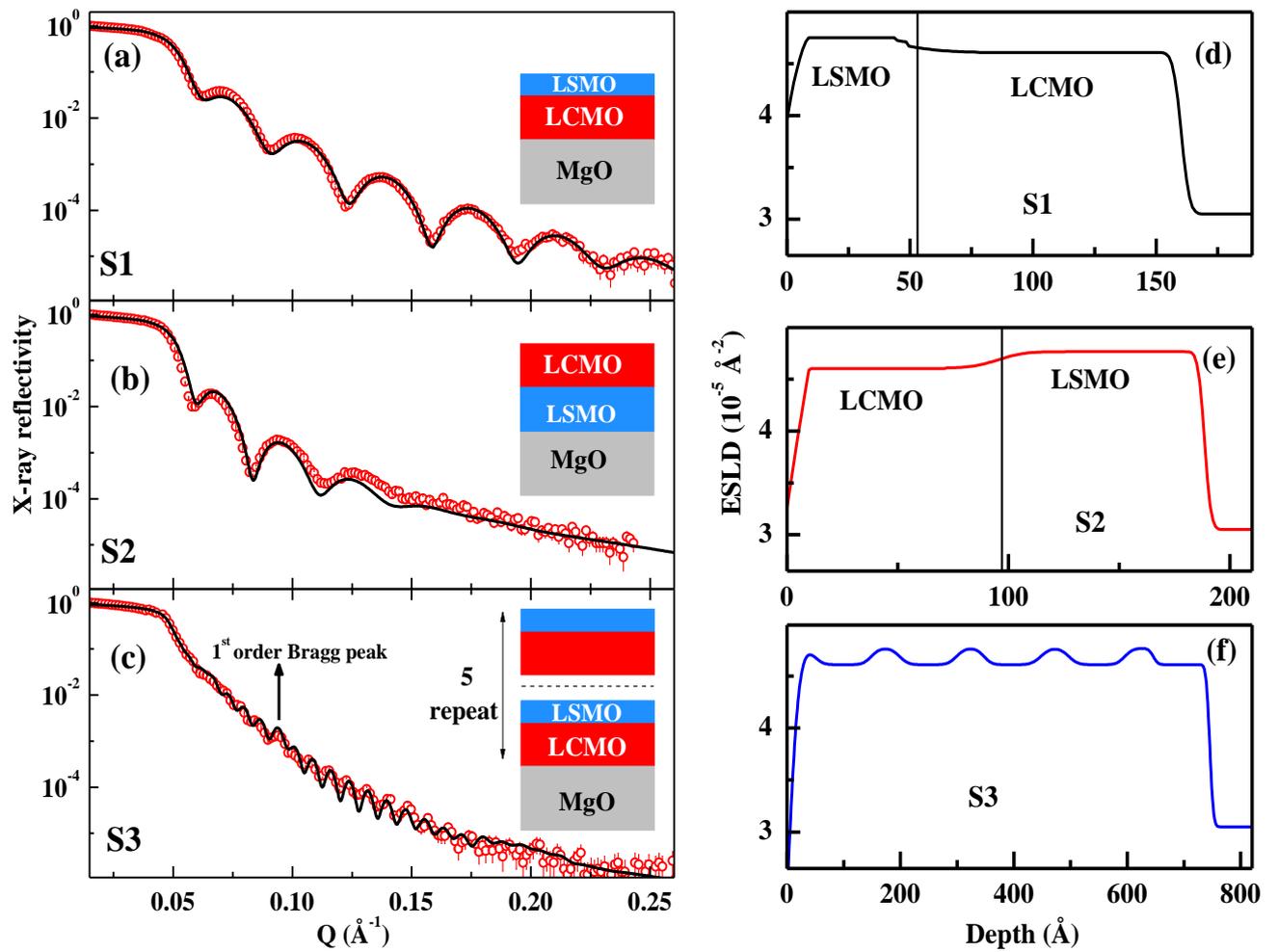

Figure S1: (a)-(c) XRR data (open circles) with corresponding best fits (solid lines) for bilayer S1, S2 and multilayer S3 heterostructures, respectively. (d)-(f) show the corresponding electron scattering length density (ESLD) depth profiles for the heterostructures S1-S3 respectively, obtained from the best fits of the experimental data.

**TEM Measurements:**

Direct imaging technique i.e. scanning transmission electron microscopy (STEM), was used to investigate the atomic scale structure along the depth of the heterostructure. Aberration-corrected scanning transmission electron microscope (JEOL-ARM) was used to acquire high-angle annular dark field (HAADF) images. In addition, energy dispersive x-ray spectroscopy (EDS) measurements were performed to distinguish the LCMO and LSMO layers. The elemental maps determined from characteristic La-$L_{\alpha 1}$, O-$K_{\alpha 1}$, Mn-$K_{\alpha 1}$, Ca-$K_{\alpha 1}$ and Sr-$K_{\alpha 1}$ EDS edges for the heterostructure S3.

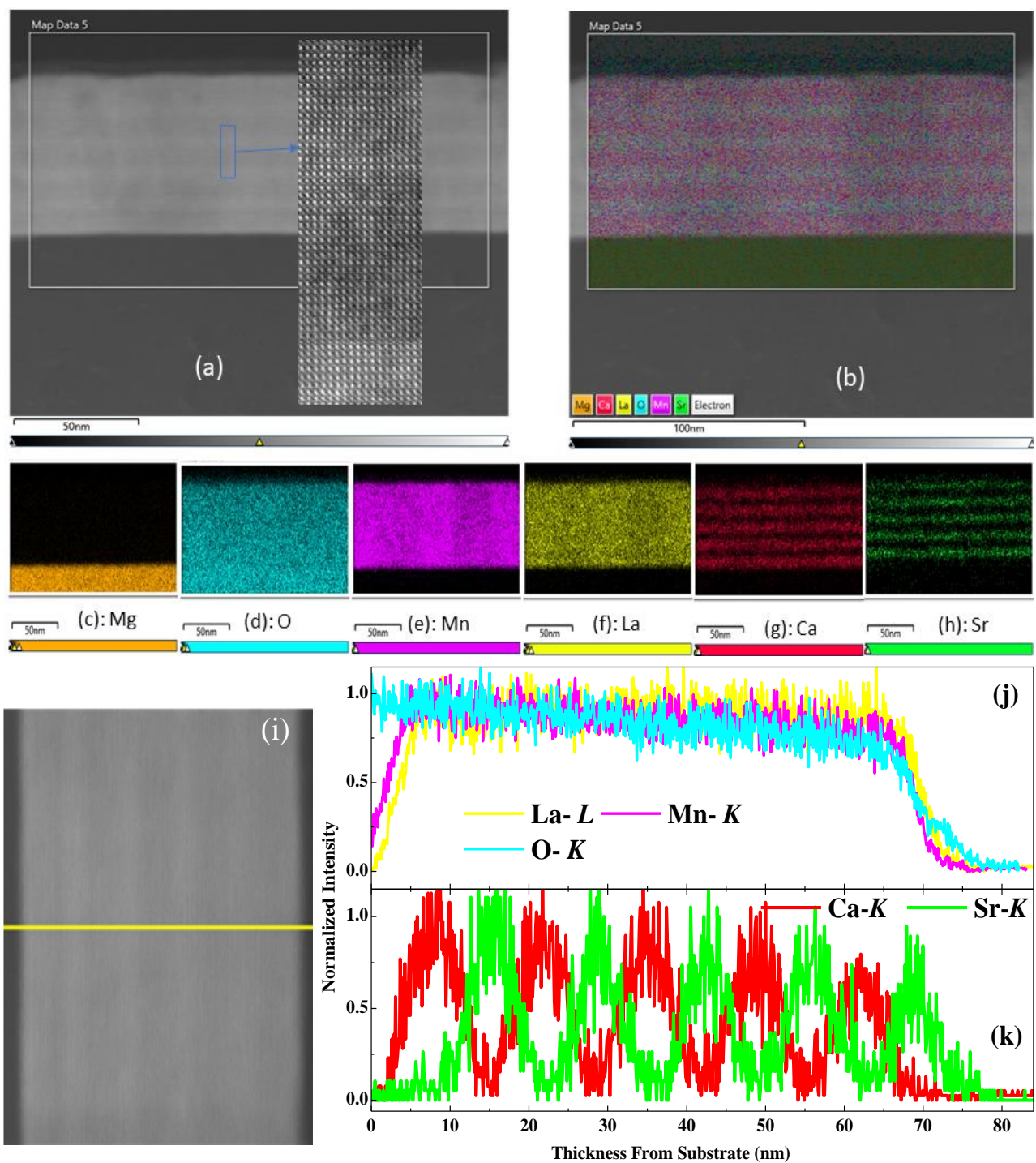

Figure S2: (a) High-angle annular dark field image of heterostructure S3 measured by high-resolution scanning transmission electron microscope (STEM). (b) elemental maps obtained by energy dispersive x-ray spectroscopy (EDS) measurements for La, O, Mn, Ca, and Sr. (c to h) show the elemental maps determined from characteristic La-$L_{\alpha 1}$, O-$K_{\alpha 1}$, Mn-$K_{\alpha 1}$, Ca-$K_{\alpha 1}$ and Sr-$K_{\alpha 1}$ EDS edges for the heterostructure S3. (i) STEM image of S3 across interfaces was scanned to measure the depth dependent elemental profiles. (j and k) the The depth profiling of elements concentration using EDS from sample S3.

**Macroscopic magnetization (SQUID) measurements:**

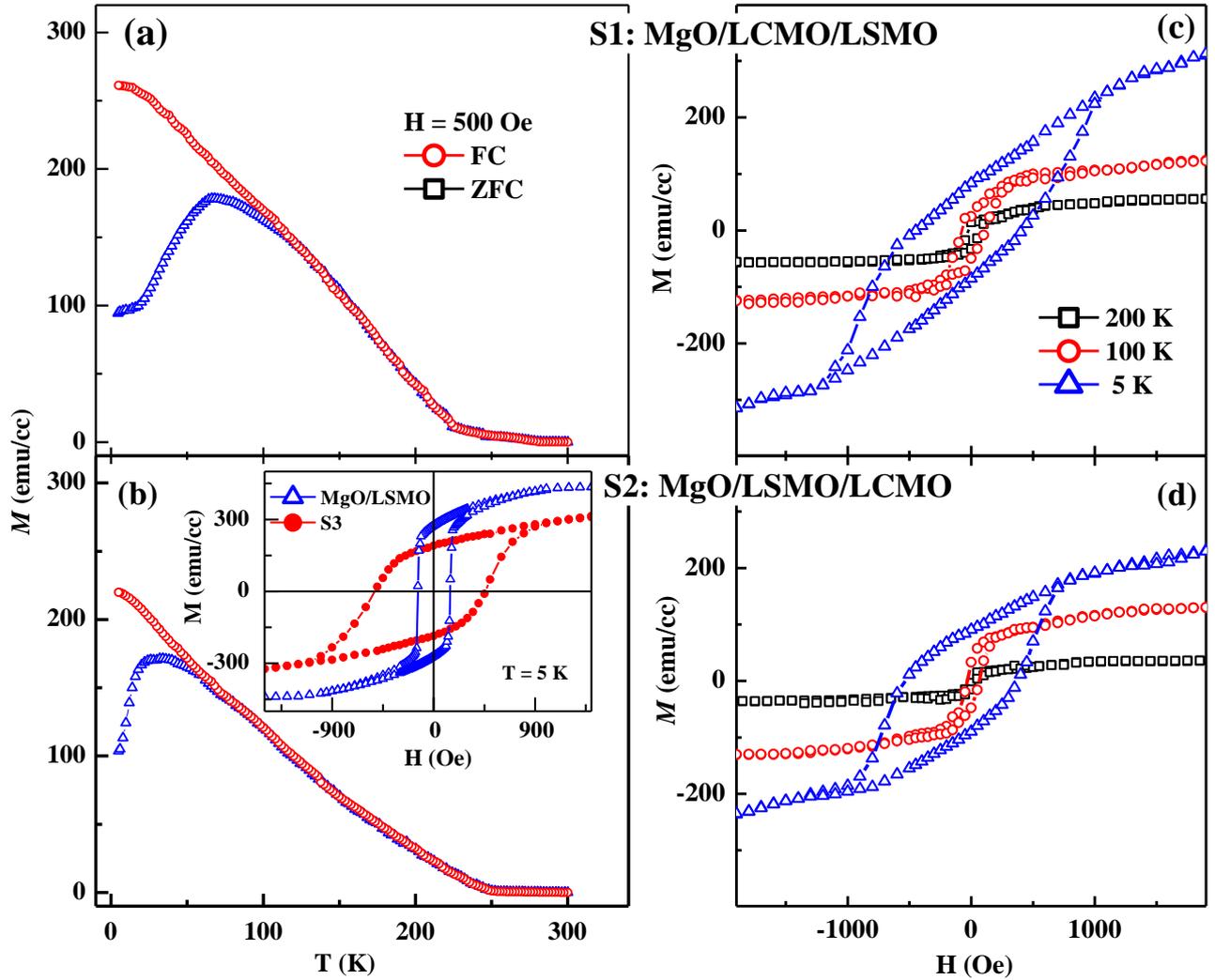

Figure S3: (a) and (b) show the *M(T)* plots under FC and ZFC conditions for S1 and S2, respectively, indicating that the $T_c$ is less than 300 K for these heterostructures. Inet of (b) shows a comparision of *M(H)* curves for a single LSMO film grown on MgO substrate (i.e. LSMO/MgO) and heterostructure S3 at 5 K, suggesting that shift in M(H) curve (Exchange bias) is observed for S3 only, which may be influenced by interface dependent magnetic structure. (c) and (d) show the *M(H)* plots at 5, 100 and 200 K for S1 and S2, respectively.

**References:**

[1]   S. Singh, M. Swain, and S. Basu, Progress in Materials Science **96**, 1 (2018).

[2]   S. Singh, M. R. Fitzsimmons, T. Lookman, J. D. Thompson, H. Jeen, A. Biswas, M. A. Roldan, and M. Varela, Physical Review Letters 108, 077207 (2012)

[3]   L. G. Parratt, Physical Review 95, 359 (1954).